\documentclass[12pt]{article}
\usepackage[dvips]{graphicx}
\usepackage{verbatim}
\usepackage[sumlimits,]{amsmath}
\usepackage{amsfonts}
\usepackage{amsthm}
\usepackage{amssymb}
\usepackage{mathrsfs}
\usepackage{srcltx}
\usepackage{bbm}
\usepackage{algorithm,algorithmic}

\usepackage{graphicx,epsfig}
\def \d { {\,\mbox{d}}}

\def \dPa { {\,\rm{dP}^a}}

\def \dt { \,\mbox{d}t}
\def \dv { \,\mbox{d}v}

\def \rmP { {\mathrm{P}}}

\def \calU { \mathcal{U}}

\def \calA {\mathcal{A}}

\def \Qsai { {Q^{sai}}}

\def \eps { {\varepsilon}}

\def \one { \mathbf{1}}

\def \pR { {\p R}}

\def \Ih { {I^h}}
\def \Is { {I^s}}
\def \pnuN { {\p_\nu N}}

\def \calU { {\mathcal{U}}}
\def \varphih { {\varphi^h}}
\def \xia { {\xi_a}}

\def \xisb { {\xi_{sb}}}
\def \xisai { {\xi_{sai}}}

\def \xih { {\xi_h}}
\def \xiheu { {\xi_{heu}}}
\def \xiq { {\xi_q}}
\newcommand{\Exp}[1]{ \E\left\{ #1 \right\} }

\newcommand{\ip}[2]{ \langle #1,\,#2 \rangle}   %Mapping is ` i 
\newcommand{\Var}[1]{ \mbox{Var}\left\{ #1 \right\} } 
\newcommand{\pfvol}[3]{ p(#1,#2\!\!\to\!\!#3)}

\newcommand{\PFsurf}[3]{ P(#1,#2\!\!\to\!\!#3)}

\def \E { {\mathbb E}}

\def \p {\partial}

\def \dsphere { {\mathbb{S}^{d-1}}}

\def \Rd { {{\mathbb R}^d}}

\def \Rone {{\mathbb R}}

\newtheorem*{theorem*}{Theorem}

\theoremstyle{definition}
\newtheorem*{def*}{Definition}

\newtheorem*{algorithmold*}{Algorithm}

\theoremstyle{remark}
\newtheorem*{remark*}{Remark}
\newtheorem{remark}{Remark}[section]

\newtheorem*{claim*}{Claim}

\begin{document}

%\title{Importance Sampling and Adjoint Hybrid Methods in Monte Carlo Transport with Reflecting Boundaries}
\title{A Hybrid (Monte-Carlo/Deterministic) Approach for Multi-Dimensional Radiation Transport}

%\author{}
\author{
Guillaume Bal\thanks{Department of Applied Physics and Applied Mathematics, Columbia University, 200 S.W. Mudd Building, 500 W. 120th Street, New York, NY, 10027, USA; +1-212-854-4731, gb2030@columbia.edu}, 
Anthony B. Davis\thanks{Jet Propulsion Laboratory, California Institute of Technology, 4800 Oak Grove Drive, Mail Stop 169-237, Pasadena, CA, 91109, USA; +1-818-354-0450, Anthony.B.Davis@jpl.nasa.gov}, and 
Ian Langmore\thanks{Corresponding author.  Department of Applied and Applied Mathematics, Columbia University, 200 S.W. Mudd Building, 500 W. 120th Street, New York, NY, 10027, USA; +1-415-272-6321, ianlangmore@gmail.com}.
}
\maketitle

\abstract{
A novel hybrid Monte Carlo transport scheme is demonstrated in a scene with solar illumination, scattering and absorbing 2D atmosphere, a textured reflecting mountain, and a small detector located in the sky (mounted on a satellite or a airplane).   It uses a deterministic approximation of an adjoint transport solution to reduce variance, computed quickly by ignoring atmospheric interactions.  This allows significant variance and computational cost reductions when the atmospheric scattering and absorption coefficient are small.  When combined with an atmospheric photon-redirection scheme, significant variance reduction (equivalently acceleration) is achieved in the presence of atmospheric interactions.
}

\vspace{12pt}


\newpage

\tableofcontents

\section{Introduction}
\label{section:introduction}

\subsection{Motivation and Background}
\label{subsection:motivation_and_background}

Forward and inverse linear transport models find applications in many areas of science including neutron transport \cite{Dav-OX-57,spanier,Lux-Koblinger}, medical imaging and optical tomography \cite{Arridge99,B-IP-09}, radiative transfer in planetary atmospheres \cite{chandra,Liou-AP-02,MD-SP-05} and in oceans \cite{Mobleyetal93,ThomasStamnes02}, as well as the propagation of seismic waves in the solid Earth \cite{sato-fehler}.  In this paper, we focus on the solution of the forward transport problem by the Monte Carlo (MC) method with, as our main application, remote sensing (an inverse transport problem) of the atmosphere/surface system \cite{Schott07}.  In our demonstration, light is emitted from the Sun and propagates in a complex environment involving absorption and scattering in the atmosphere and reflection at the Earth's surface before (a tiny fraction of) it reaches a narrowband detector, typically mounted on a airplane or a satellite.

The integro-differential transport equation \eqref{align:RTE_differential} may be solved numerically in a variety of ways.  Monte Carlo (MC) simulations model the propagation of individual photons along their path and are well adapted to the complicated geometries encountered in remote sensing.  Photons scatter and are absorbed with prescribed probability depending on the underlying medium.  The output from the simulation, e.g., the fraction of photons that hit a detector, is the expected value of a well-chosen random variable.  These simulations are very easy to code, embarrassingly parallel to run, and suffer (in principle) no discretization error. The drawback is that they can be very slow to converge.  MC methods converge at a rate $(variance/N)^{1/2}$ where $N$ is the number of simulations, and the $variance$ is that of each photon fired.  In remote sensing, the (relative) variance is high in large part because the detector is typically small and thus most photons are not recorded by the detector.  In order to be effective, even in a forward simulation, MC methods must be accelerated.

One approach to speedup MC simulations is to use quasi-Monte Carlo methods, which steepen the convergence rate from $\sim N^{-1/2}$ to a more negative exponent.  However, most MC speedup efforts focus on reducing the variance of each photon.  See \cite{spanier,Lux-Koblinger} or the review of more recent work on neutron transport in \cite{Solomon,HagWag_ProNuclEn2003_Monte,Hoogenboom} and on 3D atmospheric radiative transfer in \cite{EvansMarshak05,BurasMayer11}.  See also \cite{Veach_Thesis1997} for a thorough introduction to the MC techniques, including variance reduction, used in computer graphics.  In problems with a small detector, this is achieved by directing photons toward that detector, and re-weighting to keep calculations unbiased.  When \emph{survival-biasing} is used, photons have their weight decreased rather than being absorbed \cite{spanier,Lux-Koblinger}.\footnote{
Note the somewhat confusing terminology:  On the one hand, a method is statistically biased if the expected outcome is not the intended one.  On the other, the practice of re-directing photons in favorable directions and/or reducing the number of scattering events is also called biasing.  In the latter case the photon has its weight adjusted so that the simulation is unbiased.}  
Often, one uses some heuristic (such as proximity to the detector), or some function to measure the ``importance'' of each region of phase space.  In \emph{splitting methods} \cite{spanier,Lux-Koblinger}, the photon is split into two or more photons upon identifying that a photon is in a region of high importance.  The weight of each photon is then decreased proportionately.  Propagating many photons with a low weight is not desirable, therefore splitting is often accompanied by \emph{Russian roulette}.  Here, if a photon enters a region of low enough importance, then the photon is terminated with a certain probability, i.e., high chance of absorption if the weight is low; in the rarer alternative outcome of the Bernoulli trial, the weight is increased to keep the simulation numerically unbiased.  So there is typically a slight cost in variance to improve efficiency (by terminating low-weighted trajectories).  Typically a \emph{weight window} is used to enforce regions of low/high importance.  \emph{Source biasing} techniques change the source distribution in order to more effectively reach the detector.  More generally, the absorption and scattering properties at any point can be modified, provided photons are re-weighted correctly.

It has long been recognized that the adjoint transport solution is a natural importance function \cite{Kalos63,spanier,Lux-Koblinger,TurnerLarsen_NuclSciEng1997_Automatic1,TurnerLarsen_NuclSciEng1997_Automatic2,VanRiper_JointInterConf1997_AVATAR,
HagWag_ProNuclEn2003_Monte,DensmoreLarsen_JCompPhys2003_Variational,Hoogenboom}.  %AD: put into chronological order, here and in other lists
One can use approximations of the adjoint solution---typically a coarse deterministic solution---to reduce variance.  The result is a \emph{hybrid} method (deterministic \& MC).  The AVATAR method uses an adjoint approximation to determine weight windows \cite{VanRiper_JointInterConf1997_AVATAR}.  The CADIS scheme in \cite{HagWag_ProNuclEn2003_Monte} uses an adjoint approximation in both source biasing and weight-window determination.  An adaptive technique that successively refines the solution in ``important'' regions, using the adjoint to designate such regions, is described in \cite{Kong_Efficient_2008,Kong_ANewProof_2008}.  In \cite{Kalos63,spanier,Lux-Koblinger,TurnerLarsen_NuclSciEng1997_Automatic1}, a zero-variance technique is outlined that uses the true adjoint solution to launch photons that all reach the detector with the same weight ... which happens to be the correct answer.  This method is of course impractical since determining the exact adjoint solution everywhere is harder than determining some specific integral of that solution, which is usually the goal of a MC simulation.  The LIFT method \cite{TurnerLarsen_NuclSciEng1997_Automatic1,TurnerLarsen_NuclSciEng1997_Automatic2} therefore uses an approximation of the adjoint solution to approximate this zero-variance method.

\begin{figure}[ht!]
  \begin{center}
  \includegraphics[width=0.85\textwidth, clip=true, trim = 8cm 4cm 5cm 4.5cm]{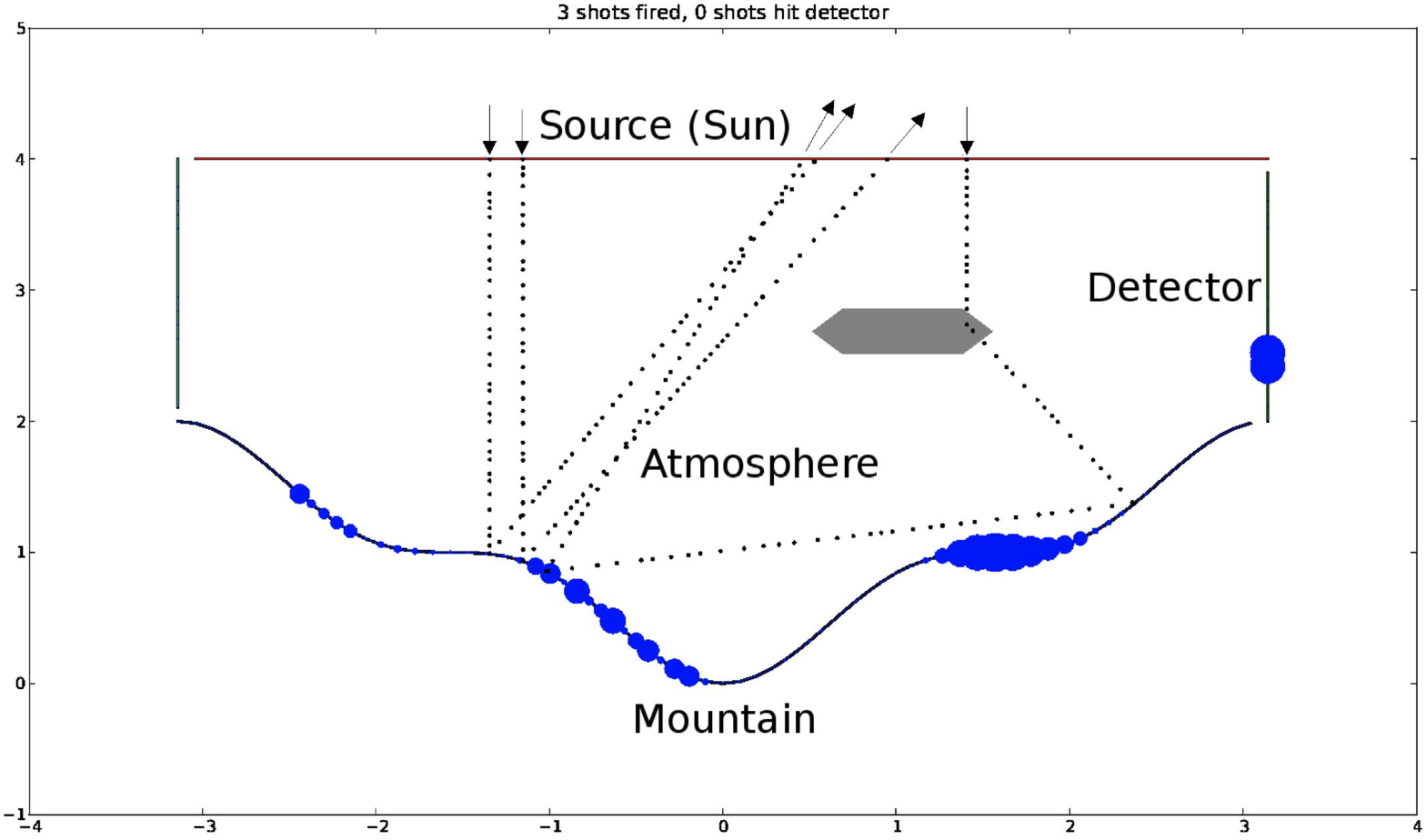}
  \caption{Mountain ($1-\cos^3x$ shape), cloud, sky, and detector.  Dot size indicates relative adjoint flux strength.  Large dots on right-hand-side are the detector (dot size is down-scaled for detector).  Dot size on mountain indicates that portions of the mountain are shaded from the detector, and that the surface albedo is varying.  See section \ref{sec:numvol} for specifics, as used in the present study.}
  \label{fig:cos3boundary}
  \end{center}
\end{figure}

We adapt the zero-variance technique to the particular problem we have at hand; see Fig.~\ref{fig:cos3boundary} for the type of geometry considered in this paper.  The problem we consider has a fixed, partially-reflective, complex-shaped lower boundary, and relatively large mean-free-path (MFP) in the sense that a large fraction of the photons reaching the detector have not scattered inside the (optically thin) atmosphere.  Calculation of the approximate adjoint solution used to emulate zero-variance techniques is difficult and potentially very costly.  What we demonstrate in this paper is that partial, ``localized'' (in an appropriate sense) knowledge of the adjoint solution still offers very significant variance reductions.  More specifically, we calculate adjoint solutions that accurately account for the presence of the boundary but do not account for atmospheric scattering (infinite MFP limit). The computation of the adjoint solution thus becomes a radiosity problem with much reduced dimensionality compared to the full transport problem.  This, of course, can only reduce variance in proportion to the number of ``ballistic'' photons that never interact with the atmosphere.  When combined with simple rules for allowing atmospheric scattering and sending some photons directly from the atmosphere to the detector, our hybrid method yields very significant variance reduction at relatively minimal cost.  Furthermore, the methodology studied is applicable whenever any method is available to deterministically pre-calculate flux over any subset of paths.  For instance, complex propagation of light in clouds and its importance could be pre-calculated locally and incorporated into the MC simulations in a similar fashion.  This ``modular'' approach to the description of the adjoint solution is well-adapted to the geometries of interest in remote sensing and avoids complicated, global (hence expensive) deterministic calculations of adjoint transport solutions.  Our treatment of the reflecting boundary described in detail in this paper is a first step toward modular adjoint transport calculations and their variance reduction capabilities in remote sensing.

The rest of the paper is organized as follows.  In sections \ref{subsection:problem_setup} and \ref{subsection:statistical_formulation} respectively, the physical problem and statistical formulation are described.  In section \ref{subsection:standard_algorithms}, the analog and survival-biased MC algorithms are presented.  In section \ref{section:sai}, the surface adjoint importance (SAI) and regularized SAI methods are introduced.  These are the hybrid adjoint-based methods at the core of this work.  In section \ref{section:numerics}, numerical estimations of variance reductions and computational speedups are given.  For an expanded exposition of techniques and analyses from a mathematical standpoint, we refer the interested reader to \cite{BalImportanceArxiv}.  Finally, we summarize our findings in section \ref{section:conclusion} and conclude with thoughts about potential applications in remote sensing science.

\subsection{Problem Setup}
\label{subsection:problem_setup}

Our setup is photon transport in a domain $R\subset\Rd$ ($d=2,3$, with $d=2$ in our present demonstration) described in Fig.~\ref{fig:cos3boundary}.  The outward normal to the domain boundary $\pR$ at position $r$ is denoted $\nu_r$.  $R$ is the atmosphere, the sky/mountain/sides/detector constitute $\pR$.  We have one small detector located on the right-side of $\pR$, and the goal of simulations is to estimate the photon flux through the detector.  Since photons are monokinetic, propagation direction $v$ is a unit vector in the sphere $\dsphere$ embedded in $d$-dimensional space (unit circle for $d=2$).

We model radiance (a.k.a. specific intensity or angular photon flux density) $I(r,v)$ in our medium with a boundary source distribution $Q$.  $I$ obeys the following integro-differential transport equation and boundary condition:
%AD: new notations above and below for \psi and S
\begin{align}
  \begin{split}
    v\cdot\nabla I(r,v) + \sigma(r)I(r,v) &= K I(r,v)\\
    I(r,v) &= \frac{K I(r,v)}{|\nu_r\cdot v|} + \frac{Q(r,v)}{|\nu_r\cdot v|},  \qquad r\in\pR,\mbox{ and } v\cdot\nu_r<0,
  \end{split}
  \label{align:RTE_differential}
\end{align}
the integral operators being defined by kernels
%AD: more new notations here for phase functions \pfvol and \PFsurf
\begin{align}
   \begin{array}{rcll}
    Kf(r,v) &=& \displaystyle \sigma_s(r)\int_\dsphere \pfvol{r}{v'}{v}f(r,v')\dv', \quad &
    r\in R \\
    Kf(r,v) &=& \alpha(r)\displaystyle\int_{\nu_{r}\cdot v'>0}
    \PFsurf{r}{v'}{v}|\nu_r\cdot v'|f(r,v')\dv'\quad & r\in \pR,\mbox{ and } v\cdot\nu_r<0.
   \end{array}
\end{align}
The extinction coefficient, a.k.a. total cross section (per unit of volume) $\sigma(r)$ is the sum of the intrinsic absorption coefficient/cross-section $\sigma_a(r)$ and the scattering coefficient/cross-section $\sigma_s(r)$.  For the partially reflecting boundary condition (viewed here as a surface scattering), $\alpha(r)$ is the local value of the albedo.  Both volume ($\pfvol{r}{v'}{v}$) and surface ($\PFsurf{r}{v'}{v}$) phase functions are normalized ($\int p\dv'=1$).

Since the transport problem is linear, we use a normalized boundary source, i.e.,
\begin{equation} 
  \int_\pR\int_{\nu_r\cdot v<0} Q(r,v) \d\mu(r) \dv = 1
  \label{eq:norms}
\end{equation} 
where $\d\mu(r)$ is the appropriate measure on the $(d-1)$-dimensional boundary.

Our detector measures photon flux and is described by a ``response function,'' $g(r,v)|\nu_r\cdot v|$, where $g(r,v)$ is zero everywhere except when $r$ is in the physical detector (its aperture or ``pupil'') and $v$ points out of the boundary.  Where $g \neq 0$ it is constant and, furthermore, it is normalized so that $\int g(r,v)\d r\dv = 1$.  The goal of our Monte Carlo method is to compute the detector's signal
\begin{align}
  \int_\pR\int_{\nu_r\cdot v>0}g(r,v)|\nu_r\cdot v|I(r,v)\d\mu(r)\dv.
  \label{align:measurement}
\end{align}
In the present study, any $v$ can contribute to the radiometric signal measured at $r$.  To model an imaging detector, direction space would be limited to a finite field-of-view that would in turn be subdivided into individual ``pixels.''

For future use, we define the function
\begin{align*}
  E_\sigma(r,r') :&= \exp\left\{ -\int_0^{|r-r'|}\sigma(r+t\widehat{r'-r})\dt \right\},
\end{align*}
where $\widehat{r'-r} := (r'-r)/|r'-r|$.  Physically, it describes the probability of \emph{direct} transmission of light from point $r$ to point $r'$ (or vice-versa), that is, without suffering any collision.

\subsection{Statistical formulation/notation}
\label{subsection:statistical_formulation}

The measurement defined formally in \eqref{align:measurement} is approximated in a Monte Carlo simulation by estimating an average 
\begin{align*}
  S_N :&= \frac{1}{N}\sum_{n=1}^N \one_D(\omega_n)
\end{align*}
where the $\omega_n$ are photon paths ($\omega=(r_0,r_1,\dots,r_k)$) generated by the ``analog'' chain (meaning analogously to real-life photon travel, cf. Algorithm \ref{alg:analog}), and the relevant indicator function is $\one_D(\omega) = 1$ if the path hits the detector (hence the subscript $D$), and $= 0$ otherwise.  The paths are random variables and, with $\Exp{\cdot}$ denoting statistical \emph{expectation}, we have 
\begin{align*}
  \Exp{S_N} &= \Exp{\one_D} = \rmP[D],
\end{align*}
where the notation $\rmP[D]$ emphasizes that this is a probability of hitting the detector.  We also have $\rmP[D]$ equal to the desired measurement or signal in \eqref{align:measurement}.
For finite $N$, $S_N$ is not equal to $\rmP[D]$ exactly.  The mismatch is quantified in a statistical sense through the \emph{variance}  
\begin{align*}
  \Var{S_N} :&= \Exp{\left( S_N-\rmP[D] \right)^2} = \frac{\Exp{\left( \one_D-\rmP[D] \right)^2}}{N} = \frac{\Var{\one_D}}{N},
\end{align*}
since all of the events $\omega_n$ contributing to the $S_N$ estimator are independently drawn.

Rather than $S_N$, one may generate paths according to some modification of real-life photon travel and then estimate
\begin{align*}
  \rmP[D] &\approx T_N := \frac{1}{N}\sum_{n=1}^N \one_D(\omega_n)\frac{\dPa\,}{\d\overline\rmP}(\omega_n),
\end{align*}
where the ratio $\dPa/\d\overline\rmP(\omega)$ is the ratio of the probability density of $\omega$ in the analog chain to that in the modified chain.  This \emph{importance sampling} technique is widely used in statistics since often times $T_N$ will have lower variance than $S_N$.  Indeed, most of the variance reduction techniques mentioned in the introduction are of this type.  In our algorithms we compute this ratio step-by-step and refer to it as a weight (modifier).  So, rather than counting photons, we count weighted photons.

For future use we define the following (standard) statistical notations and convention.  First, we write $u\sim\calU[0,1]$ to indicate that $u$ is a random variable uniformly distributed on the interval $[0,1]$.  Second, a probability density such as $\pi(x)$ can be denoted explicitly (e.g., $\pi(x) = (2\pi)^{-1/2}\exp\left\{ -x^2/2 \right\}$ in the case of the normal distribution with zero mean and unit variance), or it can be given up to a constant (since it must integrate to one).  In this last case, we would write $\pi(x) \propto \exp\left\{ -x^2/2 \right\}$.

\subsection{Standard Algorithms}
\label{subsection:standard_algorithms}

We present here two basic algorithms for Monte Carlo transport.  These are well known but we do this in order to demonstrate our notation.  Algorithm \ref{alg:analog} is often referred to as \emph{analog} since the photons follow a path analogous to photons in the real world.

\begin{algorithm}[H]
  \begin{algorithmic}[1]
    \caption{Analog Monte Carlo Transport}
    \label{alg:analog}
    \STATE Choose a starting position/direction $(r_0,v_0)$ according to the sun's source density $Q(r,v)$
    \STATE Draw $u\sim\calU[0,1]$ and cast the photon along the ray $r_0+tv_0$ until $E_\sigma(r_0+tv_0)<u$.  Call this point $r_1$.  If this does not happen before $\pR$ is reached then set $r_1$ to the boundary point at the intersection with the ray.
    \IF{$r_1\in R$}
    \STATE With probability $\sigma_s(r_1)/\sigma(r_1)$, the photon is not absorbed, and we select $v_1$ using the probability density
    \begin{align*}
      v_1&\mapsto \pfvol{r_1}{v_0}{v_1};
    \end{align*}
    otherwise the chain is stopped.
    \ELSIF{$r_1\in\pR$}
    \STATE With probability $\alpha(r_1)$ the photon is not absorbed, and we select $v_1$ using the probability density 
    \begin{align*}
      v_1&\mapsto \PFsurf{r_1}{v_0}{v_1};
    \end{align*}
    otherwise the photon is absorbed and we stop the chain.
    \ENDIF
    \STATE Continue alternating casts and direction changes until either the photon is absorbed, escapes through the upper boundary (``sky+sides''), or the detector is reached.
  \end{algorithmic}
\end{algorithm}

For use in Algorithm \ref{alg:regularized_sai} further on, we will need to know the probability density of the analog chain producing a path $\omega$.  This is given by
\begin{align*}
  D_{analog}(r_0,r_1) &= Q(r_0,v_0)E_\sigma(r_0,r_1),\\
  D_{analog}(r_0,r_1,r_2) &= D_{analog}(r_0,r_1)K^{analog}(r_1,v_0\to \widehat{r_2-r_1})E_\sigma(r_1,r_2)
\end{align*}
and so on.  Above $K^{analog}$ is given by
\begin{align*}
  K^{analog}(r_1,v_0\to v_1) :&= \left\{  
  \begin{matrix}
    \sigma_s(r_1)\pfvol{r_1}{v_0}{v_1},&\quad r_1\in R\\
    \alpha(r_1)\PFsurf{r_1}{v_0}{v_1},&\quad r_1\in\pR.
  \end{matrix}
  \right.
\end{align*}

Algorithm \ref{alg:survival_biased} uses a trick known as \emph{survival-biasing} since photons will survive (almost) any interaction with the media.  We do this by casting photons while ignoring intrinsic absorption.  So, e.g., if a patch of media has $\sigma_a$ large and $0<\sigma_s/\sigma\ll1$ the photon will almost never scatter there.  Our weight is then $E_\sigma/E_{\sigma_s}$.  When the photon interacts with the surface, then so long as $\alpha>0$, we do not absorb but multiply the photon weight by $\alpha$. Another, slightly different but also common, survival-biasing method would cast photons in the same manner as analog, but would eliminate absorption and re-weight by $\sigma_s/\sigma$.  So, e.g., if a patch of media has $\sigma_a$ large and $0<\sigma_s/\sigma\ll1$ the photon would likely interact with the media and scatter but not be absorbed there; its weight however would be reduced by a factor of $\sigma_s/\sigma$ (known in the radiative transfer literature as the ``albedo for single scattering'').  

\begin{algorithm}[H]
  \begin{algorithmic}[1]
    \caption{Survival-Biased Monte Carlo Transport}
    \label{alg:survival_biased}
    \STATE Choose a starting position/direction $(r_0,v_0)$ according to the source density $Q(r,v)$
    \STATE Draw $u\sim\calU[0,1]$ and cast the photon along the ray $r_0+tv_0$ until $E_{\sigma_s}(r_0+tv_0)<u$.  Call this point $r_1$.  If this does not happen before $\pR$ is reached then $r_1$ is the boundary point we have reached.
        Since we paid no attention to intrinsic absorption during the cast, the photon picks up a weight equal to
        \begin{align*}
          E_{\sigma_a}(r_0,r_1) &= \frac{E_\sigma(r_0,r_1)}{E_{\sigma_s(r_0,r_1)}}
        \end{align*}
    \IF{$r_1\in R$}
    \STATE Select $v_1$ using the probability density
    \begin{align*}
      v_1&\mapsto \pfvol{r_1}{v_0}{v_1}.
    \end{align*}
    \ELSIF{$r_1\in\pR$ and $\alpha(r_1)>0$}
    \STATE Select $v_1$ using the probability density 
    \begin{align*}
      v_1&\mapsto \PFsurf{r_1}{v_0}{v_1}.
    \end{align*}
    Since we had no chance of boundary absorption, the photon's weight is multiplied by $\alpha(r_1)$.
    \ELSIF{$r_1\in\pR$ and $\alpha(r_1)=0$}
    \STATE The photon is absorbed and we stop the chain.  
    \ENDIF
    \STATE Continue alternating casts and direction changes until either the photon is absorbed, escapes, or reaches the detector.
  \end{algorithmic}
\end{algorithm}

\section{The Surface Adjoint Importance (SAI) Method}
\label{section:sai}

The SAI method uses an approximation to the surface reflection problem to reduce variance coming from surface interactions.  It ignores atmospheric effects and therefore, by itself, is statistically biased.  In section \ref{subsection:regularized_sai} we pair it with other methods to produce an unbiased estimate of the detected flux.

\subsection{Pure SAI}
\label{subsection:pure_sai}

Here we ignore atmospheric effects and demonstrate and develop a Monte Carlo method that sends photons from surface point to surface point and then to the detector.  If atmospheric effects are not present, and our deterministic solution was perfectly accurate, this method would have zero variance.

The adjoint solution to transport may be developed by considering the $L^2$ adjoint of the integral solution to transport and reversing the role of the source and detector.  
Let $\Is$ be the adjoint solution when only surface effects are present.  We therefore have
\begin{align}
  \Is(r,v) &= \alpha(r)\int_{\nu_r\cdot v'<0}\PFsurf{r}{v}{v'}\Is(r_+(r,v'),v')\dv' + g(r,v).
  \label{eq:psis}
\end{align}
This adjoint solution corresponds (in a Monte Carlo viewpoint) to sending photons that start at the detector and travel backwards.  Therefore, 
it will have its maximum at the detector.  It will be higher in places that have a clear path to the detector.  $\Is$ will be zero at places from which a photon cannot reach the detector.  Our numerical solution of \eqref{eq:psis} is described in the Appendix.

The pure SAI chain is defined by the steps described in the following Algorithm~\ref{alg:pure_sai}.

\begin{algorithm}[H]
  \begin{algorithmic}[1]
    \caption{Pure SAI}
    \label{alg:pure_sai}
    \STATE Choose a starting position/direction $(r_0,v_0)$ according to the modified source density
    \begin{align*}
     \Qsai(r,v) &\propto Q(r,v)\Is(r_+(r,v),v).
    \end{align*}
    The photon picks up a weight $Q(r_0,v_0)/\Qsai(r_0,v_0)$
    \STATE Cast the photon until it hits the opposing boundary at point/direction $(r_1,v_1)=( r_+(r_0,v_1),v_1)$.  The weight is multiplied by
    \begin{align*}
      \frac{E_\sigma(r_0,r_1)}{1}.
    \end{align*}
    \STATE Change direction according to the density
    \begin{align*}
      K^{sai}(r_1,v_1\to v_2)&=C_{sai}(r_1)\PFsurf{r_1}{v_1}{v_2}\frac{\Is(r_+(r_1,v_2),v_2)}{\Is(r_1,v_1)}
    \end{align*}
    where $C_{sai}(r)$ is a normalization factor depending only on $r\in\pR$.   Since we did not account for boundary absorption, the photon weight is multiplied by $\alpha(r_1)$.  The modified direction change must also be taken into account and therefore, in addition, the weight is multiplied by 
    \begin{align*}
      \frac{K^{analog}(r_1,v_1\to v_2)}{K^{sai}(r_1, v_1\to v_2)}.
    \end{align*}
    \STATE Cast the photon until it hits the opposing boundary.  If it hits the detector, stop and record a hit.  Else, repeat step 3.
  \end{algorithmic}
\end{algorithm}
For use in Algorithm \ref{alg:regularized_sai}, we will need to compute the probability density of a path $\omega=(r_0,\dots,r_\tau)$ being generated by pure SAI.  This is simply the denominator in the corresponding weights.  Denoting this by $D_{sai}$ we have
\begin{align*}
  D_{sai}(\omega) &= 0,\qquad \mbox{if $r_j\in R$ for any $j$},
\end{align*}
and for paths such that $r_j\in\pR$ for all $j$, we define $D_{sai}$ recursively (with $v_j:= \widehat{r_{j+1}-r_j}$)
\begin{align*}
  D_{sai}(r_0,r_1) &= S^{sai}(r_0,v_0) \\
  D_{sai}(r_0,r_1,r_2) &= D_{sai}(r_0,r_1)K^{sai}(r_1,v_0\to v_1),\\
  D_{sai}(r_0,\dots,r_k) &= D_{sai}(r_0,\dots,r_{k-1})K^{sai}(r_{k-1},v_{k-2}\to v_{k-1}).
\end{align*}

\begin{remark} 
  \label{remark:pure_sai}
  \mbox{}\\
  \begin{itemize}
    \item A discretized version of the density $\Qsai$ is pre-computed using the (discrete) solution $\Is$; see section \ref{section:numerics}.  This means that we can pre-compute the normalization factor $C_{sai}$.  The discrete density $\Qsai$ will be defined at a number of points $(r_i,v_0)$ where $v_0$ is the anti-solar direction.  We use the density to decide on a center point $r_i$, and then perturb the starting point by a small (random) amount to eliminate discretization effects in the final solution.
    \item The direction change pdf is also pre-computed and stored as a discrete pdf over angles.  We use the pdf to pick a direction center $v_j$ and then perturb to obtain the new direction.
    \item Up to numerical error one can see that $C_{sai}(r) = \alpha(r)$.  Indeed, dividing \eqref{eq:psis} through by $\Is(r,v)$ we have
      \begin{align*}
        1 &= \alpha(r)\int_{\nu_r\cdot v'<0}\PFsurf{r}{v}{v'}\frac{\Is(r_+(r,v'),v')}{\Is(r,v)}\dv' + \frac{g(r,v)}{\Is(r,v)}.
      \end{align*}
      So away from the detector $g(r,v)=0$ and the integral is therefore equal to $1$.  
    \item That this method is biased is easy to see:  If a region of the atmosphere has non-zero scattering, then it would be possible (in the analog world) to scatter from that point to the detector.  This type of interaction is not allowed in a pure SAI world.
  \end{itemize}
\end{remark}

This is an implementation of the zero variance adjoint-based chains studied in \cite{spanier,Lux-Koblinger,TurnerLarsen_NuclSciEng1997_Automatic1} in the special case where atmospheric effects are not present.  Hence (disregarding numerical error), this would be a zero-variance method were atmospheric absorption/scattering absent.

\begin{figure}[ht!]
  \begin{center}
  \includegraphics[width=0.43\textwidth]{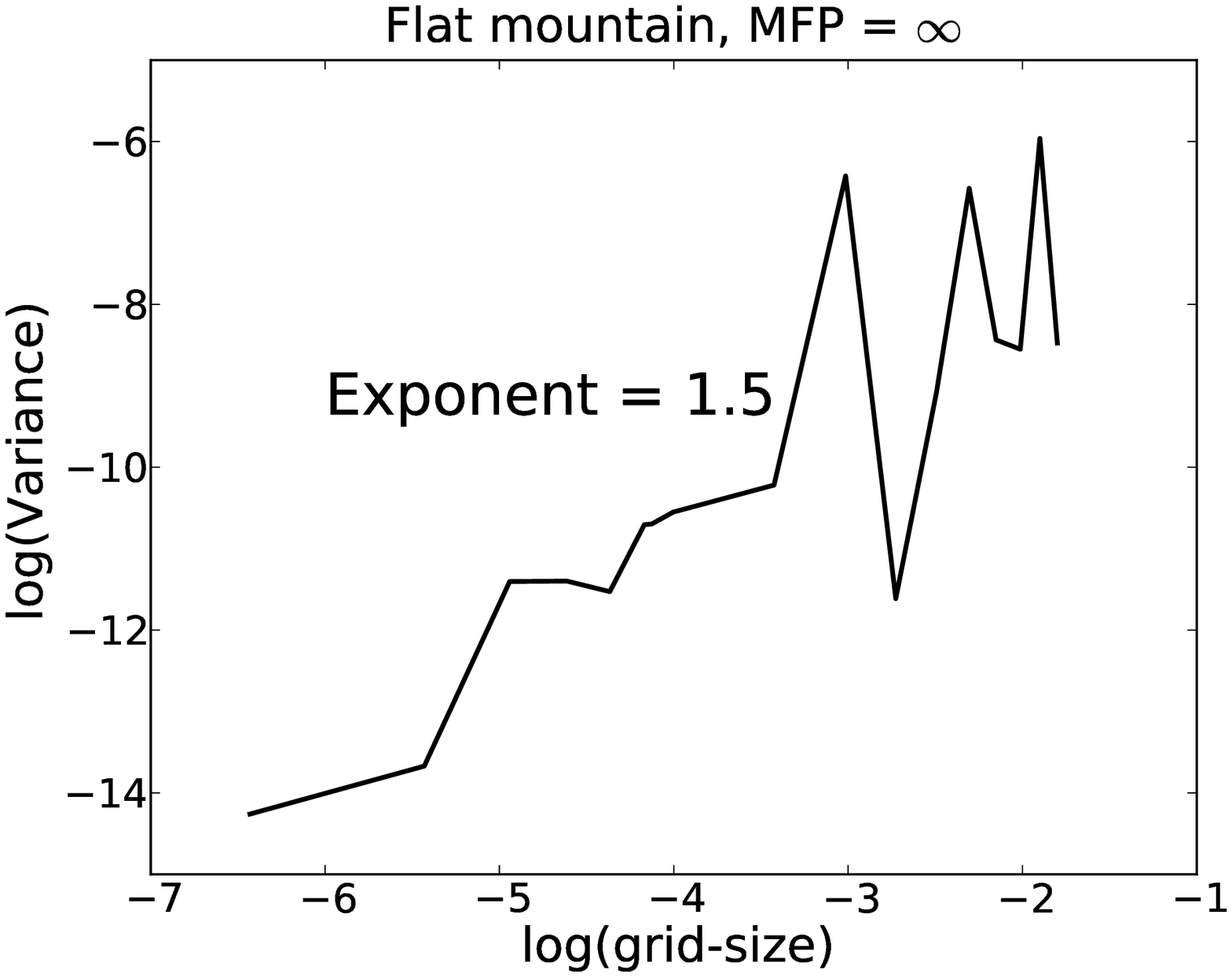}
  \includegraphics[width=0.43\textwidth]{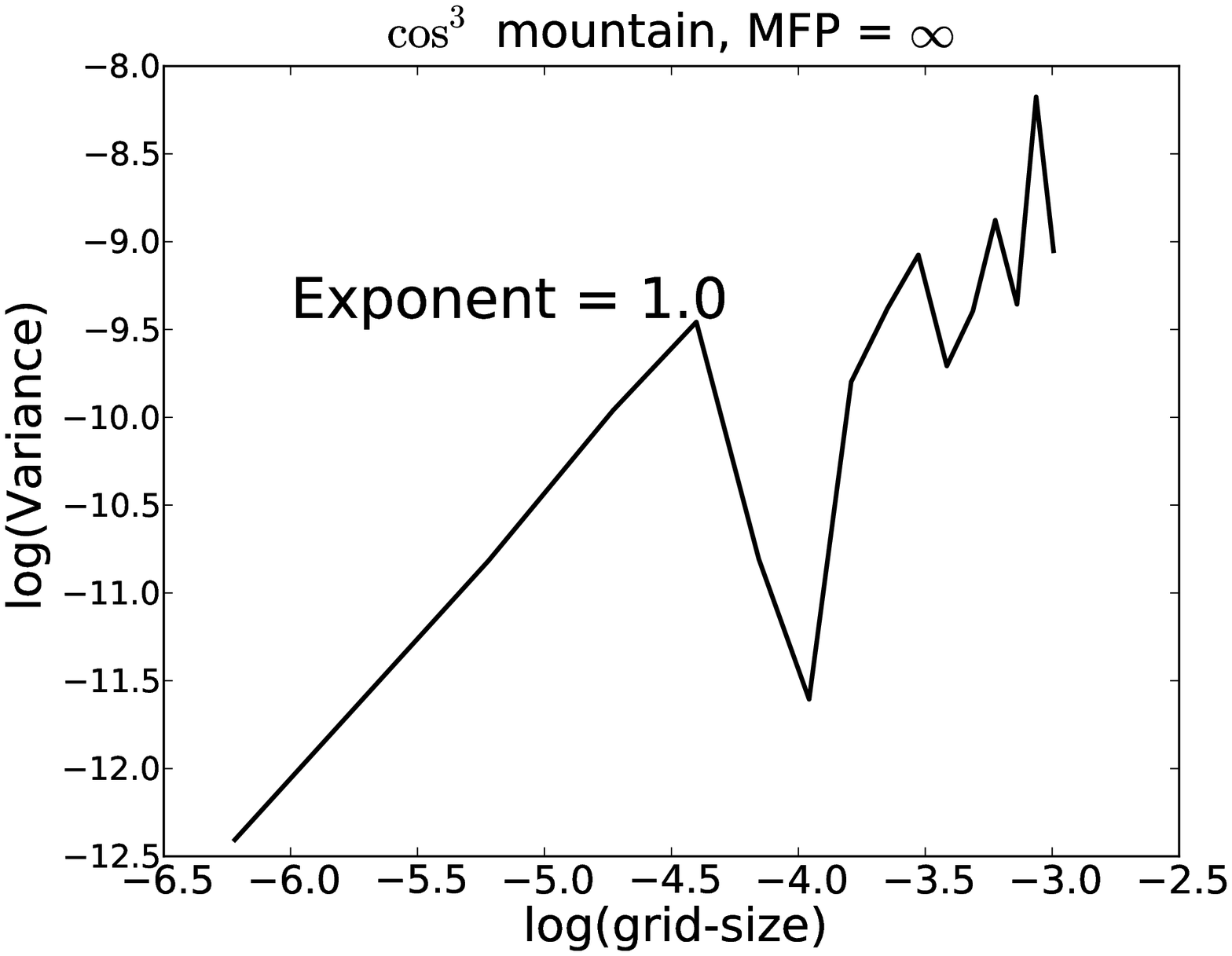}
  \caption{Left:  When a flat mountain is used, variance $\sim O(h^{1.5})$ where $h$ is the discretization parameter.  Right: On the curved boundary discretization effects are more prevalent and convergence is slower.}
  \label{figure:MFPinfinity_var_h}
\end{center}
\end{figure}

We verify this claim numerically by testing the method in simulations without atmospheric effects.  See Fig.~\ref{figure:MFPinfinity_var_h} where this is tested with both a flat terrain and a curved ``$\cos^3$'' mountain.  The curved mountain increases variance since discretization does not allow the function $r_+(x,v)$ to be implemented perfectly.

\subsection{Regularized SAI}
\label{subsection:regularized_sai}

Here we use the SAI chain as part of a larger unbiased chain. Since Algorithm \ref{alg:pure_sai} does not generate paths following all possible interactions, we must supplement it with an algorithm that does. We then use a number $q_s\in[0,1]$ to determine the fraction of photons that travel according to Algorithm \ref{alg:pure_sai} (this fraction $=1-q_s$), and what fraction according to the supplemental algorithm.

Before describing the regularized SAI algorithm, we present the supplemental Algorithm \ref{alg:heuristic} dubbed ``heuristic scattering adjustment.''  It is a survival-biased algorithm in the sense that no absorption occurs within the atmosphere, or at boundary points (unless the boundary point had $\alpha=0$, e.g. the sides/sky).  It also makes use of a simple scheme to direct a fraction of atmospheric interactions toward the detector.  No claim is made to the optimality of this re-direction (it is similar to the technique of \emph{local estimation} \cite{Marchuk_etal80,EvansMarshak05}).  We use Algorithm \ref{alg:heuristic} since it is simple to understand and illustrates the dramatic decrease in variance that can be achieved when two methods (SAI and heuristic) are used together in Algorithm \ref{alg:regularized_sai} (see also section \ref{section:numerics}).  

\begin{algorithm}[H]
  \begin{algorithmic}[1]
    \caption{Heuristic Scattering Adjustment with Parameter $q_v\in[0,1]$}
    \label{alg:heuristic}
    \STATE Choose a starting position/direction according to the standard source density $Q(r,v)$
    \STATE Cast the photon as in Algorithm \ref{alg:survival_biased} until it hits the opposing boundary or interacts with the atmosphere at $r_1$.  The photon picks up a weight equal to $E_\sigma(r_0,r_1)/E_{\sigma_s}(r_0,r_1)$
    \IF{$r_1 \in R$}
    \STATE With $r_{d_0}=$``the midpoint of the detector'', compute 
    \begin{align*}
      q_{heu}(r_1,v_0) :&= 1 - (1-q_v)\frac{\pfvol{r_1}{v_0}{\widehat{r_{d_0}-r_1}}}{\|\pfvol{r_1}{v_1}{\cdot}\|_{L^\infty}}
    \end{align*}
    With probability $1-q_{heu}$ draw $v_1$ from a uniform distribution of directions pointed toward the detector (we call this $f_V(r_1,v_1)$), and with probability $q_{heu}$ draw $v_1$ from $\pfvol{r_1}{v_0}{\cdot}$.  The weight is multiplied by
    \begin{align*}
      \frac{\sigma_s(r_1)\pfvol{r_1}{v_0}{v_1}}{(1-q_{heu})f_V(r_1,v_1) + q_{heu}\pfvol{r_1}{v_0}{v_1}},&\qquad\mbox{if $r_1\in R$}.
    \end{align*}
    \ELSIF{$r_1\in \pR$ and $\alpha(r_1)>0$} 
    \STATE pick a new direction according to the density $\PFsurf{r_1}{v_0}{v_1}$.  The weight is multiplied by $\alpha(r_1)$.
    \ELSIF {If $r_1\in\pR$ and $\alpha(r_1)=0$}  
    \STATE the photon is absorbed and we stop.  
    \ENDIF
    \STATE Continue in this manner until absorption or the detector is reached
  \end{algorithmic}
\end{algorithm}

So at every scattering event, the weight is modified by a ratio of either $\alpha(r) P$ or $\sigma_s(r) p$ to $K^{heu}$ where
\begin{align*}
  K^{heu}(r,v\to v') :&= \left\{ 
  \begin{matrix}
    (1-q_{heu})f_V(r,v) + q_{heu}\pfvol{r}{v}{v'},&\qquad  r\in R\\
    \PFsurf{r}{v}{v'},&\qquad r\in \pR.
  \end{matrix}
  \right.
\end{align*}
For use in Algorithm \ref{alg:regularized_sai} we will need to compute the probability density of a given path generated by Algorithm \ref{alg:heuristic}.  This is simply the denominator in the corresponding weight.  Denote this by $D_{heu}(r_0,r_1,\dots,r_k)$, which we define recursively by (with $v_j:=\widehat{r_{j+1}-r_j}$)
\begin{align}
  \label{align:heuristic_density}
  \begin{split}
    D_{heu}(r_0,r_1) &= Q(r_0,v_0)E_{\sigma_s}(r_0,r_1),\\
    D_{heu}(r_0,r_1,r_2) &= D_{heu}(r_0,r_1)K^{heu}(r_1,v_0\to v_1)E_{\sigma_s}(r_1,r_2),\\
    D_{heu}(r_0,\dots,r_k) &= D_{heu}(r_0,\dots,r_{k-1})K^{heu}(r_{k-1},v_{k-2}\to v_{k-1})E_{\sigma_s}(r_{k-1},r_k),
  \end{split}
\end{align}
and so on. 

We now present Algorithm \ref{alg:regularized_sai}, the regularized SAI algorithm that combines pure SAI (Algorithm \ref{alg:pure_sai}) with the heuristic scattering adjustment (Algorithm \ref{alg:heuristic}).  Note that any unbiased algorithm may be combined with pure SAI in a similar manner.

\begin{algorithm}[H]
  \begin{algorithmic}[1]
    \caption{Regularized SAI with parameters $q_s,q_v\in[0,1]$}
    \label{alg:regularized_sai}
    \STATE With probability $1-q_s$, generate a path according to Algorithm \ref{alg:pure_sai}.  With probability $q_s$ generate it according to Algorithm \ref{alg:heuristic}.
    \STATE The weight of the path $\omega=(r_1,\dots,r_\tau)$ is
    \begin{align*}
      \frac{D_{analog}(\omega)}{(1-q_s)D_{sai}(\omega) + q_sD_{heu}(\omega)}.
    \end{align*}
  \end{algorithmic}
\end{algorithm}

Algorithm \ref{alg:regularized_sai} uses SAI to produce paths that interact only with the surface.  One could easily devise other algorithms that send paths via the heuristic chain, and once paths interact with the surface they use the SAI chain.  This could reduce variance further, but we choose not to study this in order to simplify the presentation.

\section{Numerical Results}
\label{section:numerics}

%Here we present numerical results.  

\subsection{Parameter choices in numerical simulations}
\label{sec:numvol}

In the assumed $d=2$ transport space, we have $r=(x,y)$, where $x$ increases from left to right in Fig.~\ref{fig:cos3boundary} and $y$ increases from bottom to top; $r=(0,0)$ is the point at the bottom of the valley.  For directions, we have $v=v(\phi)=(\cos\phi,\sin\phi)$ where $\phi$ increases counterclockwise from the $x>0$ axis.

In the simulations performed with $\sigma=0$ (no atmospheric interactions), we used both a flat surface (so that our domain was $[-\pi,\pi]\times[2,4]$) and a ``$\cos^3$'' surface (Fig.~\ref{fig:cos3boundary}).  We swept $h$, with $0.002<h<0.2$.  We did not use any heuristic scattering adjustment ($q_v=1.0$).  In all cases, we assume an isotropic (Lambertian) redistribution by diffuse surface reflection.  This leads to the following surface scattering phase function and assumed surface albedo distribution:
\begin{align*}
  \PFsurf{r}{v}{v'} &\propto \left\{  
  \begin{matrix}
    |\nu_r\cdot v'|,& \nu_r\cdot v'<0 \\
    0,&\mbox{ otherwise}
  \end{matrix}
  \right. ;\qquad
  \alpha(r)= \left\{
  \begin{matrix}
    1,& |x|<2.5\\
    0 ,& \mbox{ otherwise}
  \end{matrix}
  \right. .
\end{align*}
The cutoff $|x|<2.5$ was done to simplify the coding (allowed us to use one simple routine for all values of $h$), and has no theoretical consequence.  The source was mono-directional $\phi=-\pi/2$, and given by
\begin{align*}
  Q(r,v(-\pi/2)) &= \left\{
  \begin{matrix}
    1/5\quad& |x|<2.5,\, y=4\\
    0\quad& \mbox{ otherwise.}
  \end{matrix}
  \right.
\end{align*}

In the simulations involving atmospheric interactions ($\sigma>0$), we used a $\cos^3$ type surface.  We compute speedup in a variety of cases. The mean-free-path MFP$=\sigma^{-1}$ was varied as well as $q_s$, $h$, and $q_v$.  We swept $0.002<h<0.15$.  In all cases the atmospheric scattering coefficients were constant with $\sigma_s=2\sigma_a$ (hence $\sigma_s/\sigma = 2/3$).  The atmospheric scattering was given by
\begin{align*}
  \pfvol{r}{v}{v'} &\propto 1 + (v\cdot v')^2,
\end{align*}
which mimics a molecular (Rayleigh) in $d=2$.  The other coefficients were chosen to have features (in this case oscillations) on a scale coarser than the fine values of $h$, and finer than the coarse values.

The surface albedo was chosen to be quite complex (significantly different than the flat surface/constant reflection commonly used).  The phase function $P$ was as before (Lambertian), but $\alpha$ is given by
\begin{align*}
  \alpha(r)&=\left\{
  \begin{matrix}
    0& |x|>2.5,\\
    0.75 + 0.25\sin(2\pi x/0.05)& 1<x<2.5,\\
    0.35 + 0.25\sin(2\pi x/0.05)& -2.5<x<1,
  \end{matrix}
  \right.
\end{align*}
using the same inconsequential cutoff $|x|<2.5$.  Off the mountain there was no scattering (perfectly absorbing boundary).

The source was mono-directional $\phi=-\pi/2$ and given by
\begin{align*}
  Q(r,v(-\pi/2)) &\propto \left\{
  \begin{matrix}
    1 + 0.25\sin(2\pi x/0.07)\quad & |x|<2.5,\, y=4\\
    0& \mbox{ otherwise}
  \end{matrix}
  \right. .
\end{align*}

\subsection{Speedup (figure of merit)}
\label{subsection:fom}

We start by defining our figure of merit used to compare the different algorithms.  We take the viewpoint that each algorithm produces a sequence of paths $\left\{ \omega^n \right\}_{n=1}^N$ and corresponding random variables $\xi(\omega^n)$ equal to the product of $\one_D(\omega^n)$ times the weight that the photon picked up along the way.  To distinguish different methods we write $\xia$ for analog, $\xisb$ for survival-biasing, $\xisai$ for pure SAI, $\xiheu$ for heuristic scattering, and $\xiq$ for the regularized SAI method.

For all of these methods, define the approximation after $N$ random draws 
\begin{align*}
  I_N(\xi) :&= \frac{1}{N}\sum_{n=1}^N \xi(\omega_n).
\end{align*}
For $\xi$ equal to any of the above methods, $I_N(\xi)$ is an unbiased estimator of $\Exp{\xi}=\rmP[D]$, i.e., the probability of a detector hit.

The RMS estimation error $\varepsilon$ is given by
\begin{align*}
  \varepsilon(\xi) :&= \sqrt{\Exp{|I_N(\xi)-\rmP[D]|^2}} = \sqrt{\frac{\Var{\xi}}{N}}.
\end{align*}
For a given error level $\varepsilon$, the required number of MC draws is then $N(\eps,\xi) := \Var{\xi}/\eps$.
The required simulation time $T(\varepsilon,\xi)$ for one estimation of $\rmP[D]$ is given by
\begin{align*}
  T(\varepsilon,\xi) :&= T_0(\xi) + \tau(\xi)N = T_0(\xi) + \frac{\tau(\xi)\Var{\xi}}{\varepsilon^2},
\end{align*}
where $T_0(\xi)$ is the time needed to compute the deterministic adjoint solution (e.g. at level $h$ when $\xi=\xih$), and $\tau(\xi)$ is the expected time for one draw using the appropriate measure for the random variable $\xi$.  We foresee the use of SAI in situations where the boundary remains fixed, but the atmosphere changes (due to, e.g., moving clouds over a fixed surface).  We therefore consider the time for $m$ simulations using one boundary,
\begin{align*}
  T(\varepsilon,\xi, m) :&= T_0(\xi) + m\tau(\xi)N = T_0(\xi) +m\frac{\tau(\xi)\Var{\xi}}{\varepsilon^2},
\end{align*}

Schemes may be compared with the ratio
\begin{align*}
  \frac{T(\varepsilon,\xi_1,m)}{T(\varepsilon,\xi_2,m)} = \frac{\eps^2T_0(\xi_1) + m\tau(\xi_1)\Var{\xi_1}}{\eps^2T_0(\xi_2) + m\tau(\xi_2)\Var{\xi_2}}.
\end{align*}

For a deterministic approximation of $\Is$, we expect $T_0(\xi)\approx C(\xi)h^{-2(d-1)}$.  We in fact measure (with $d=2$) $T_0(\xih)\approx 0.017h^{-2}$.  Our ``benchmark'' scheme is survival-biasing.  Since $\xisb$ requires no deterministic solution, the relevant ratio (and our figure of merit) is
\begin{align*}
  \mbox{Speedup}(\xiq,\eps,m) :&= \frac{m\tau(\xisb)\Var{\xisb}}{\left( \frac{\eps}{h} \right)^2C  + m\tau(\xiq)\Var{\xiq}}.
\end{align*}

We measured speedup when either $m=10$ or, formally, $m=\infty$ (``Ignoring deterministic solve'').  

\subsection{Variance reduction}
\label{subsection:varred}

\begin{figure}[h!]
    \begin{center}
  \includegraphics[width=0.43\textwidth]{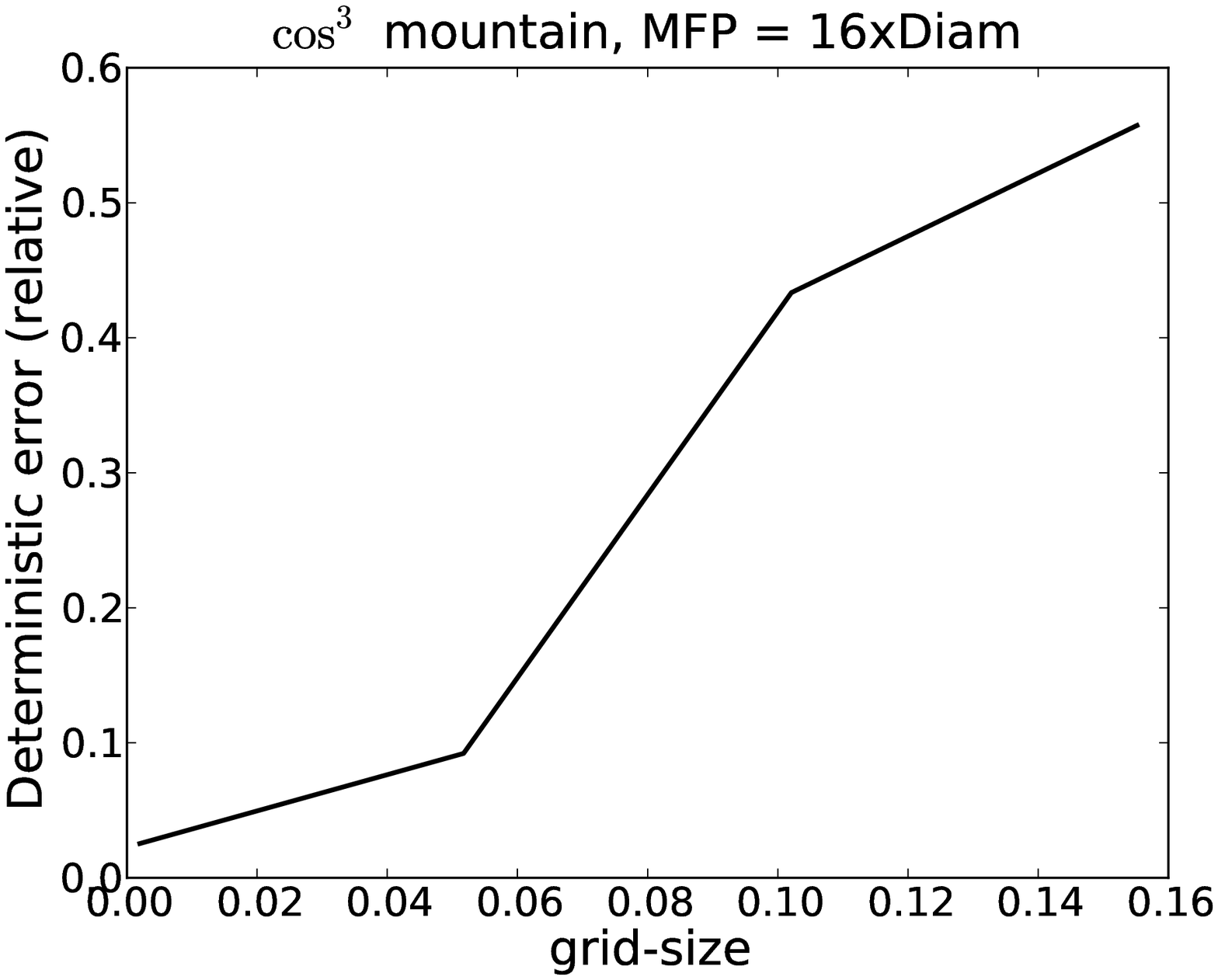}
  \includegraphics[width=0.43\textwidth]{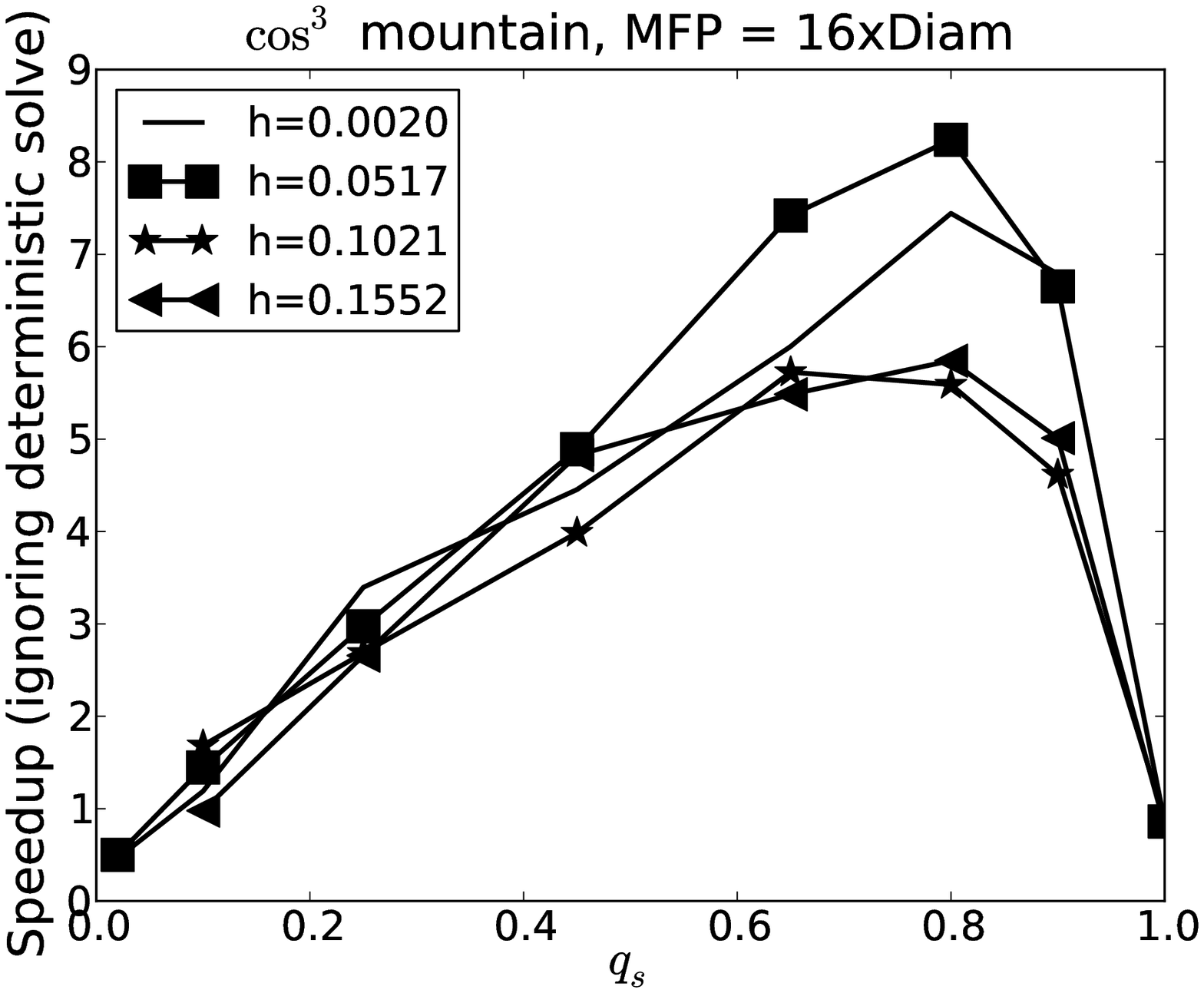}
  \caption{$|\rmP[D] - \ip{\Is}{S}|/\rmP[D]$ is generally lower for smaller $h$.  However, speedup is still very good even for large $h$.  Diam is the maximal diameter of the simulation domain $R$}
  \label{figure:deterministic_error}
  \end{center}
\end{figure}

Here we analyze the variance of the SAI chain in the presence of atmospheric interactions.
Note that even when the error $|\rmP[D] - \ip{\Is}{S}|$ is high, we still get good variance reduction.  See Fig.~\ref{figure:deterministic_error}.  This emphasizes the point that the quality of the deterministic solve is not so important in a modular scheme.

\begin{figure}[h!]
    \begin{center}
  \includegraphics[width=0.43\textwidth]{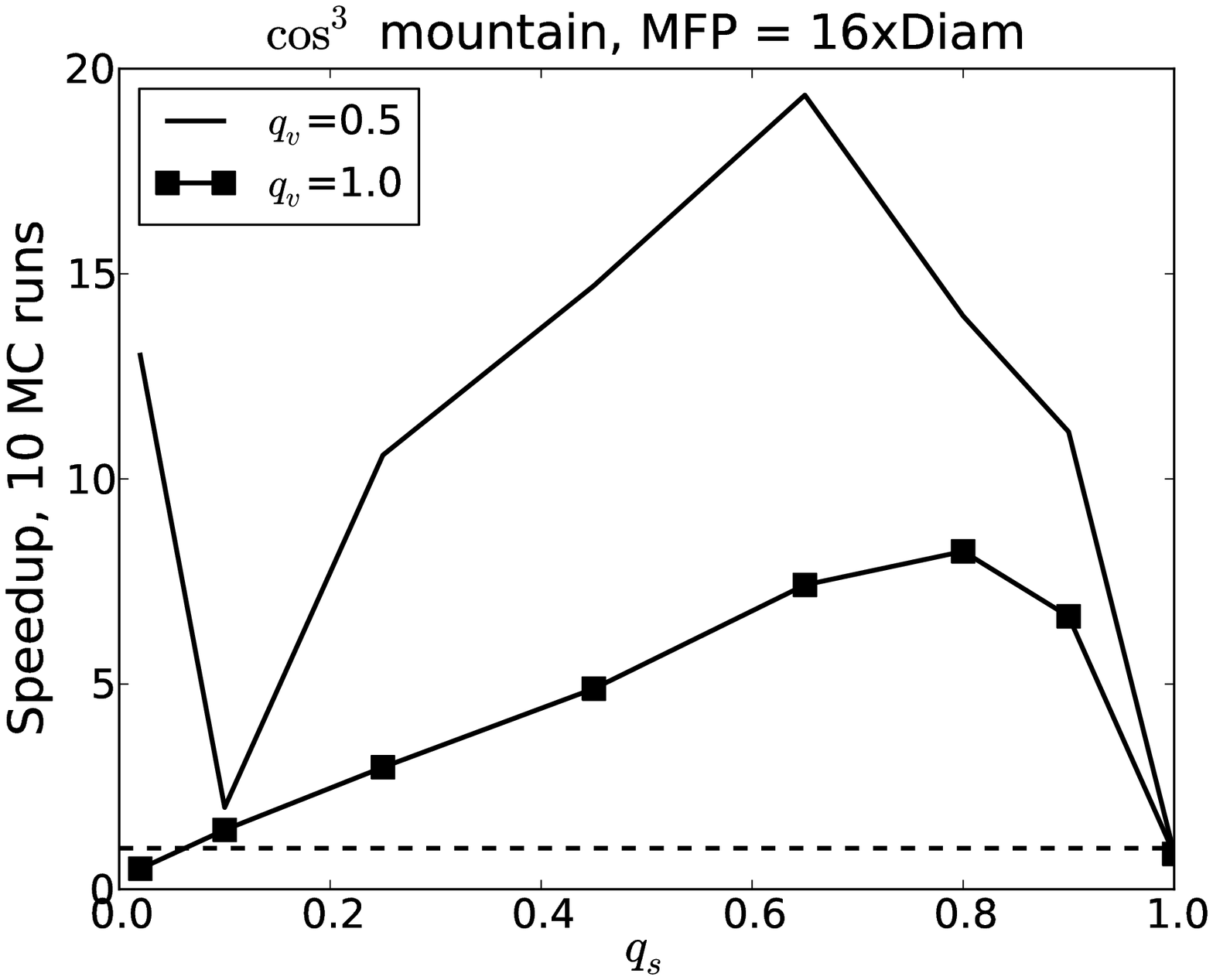}
  \includegraphics[width=0.43\textwidth]{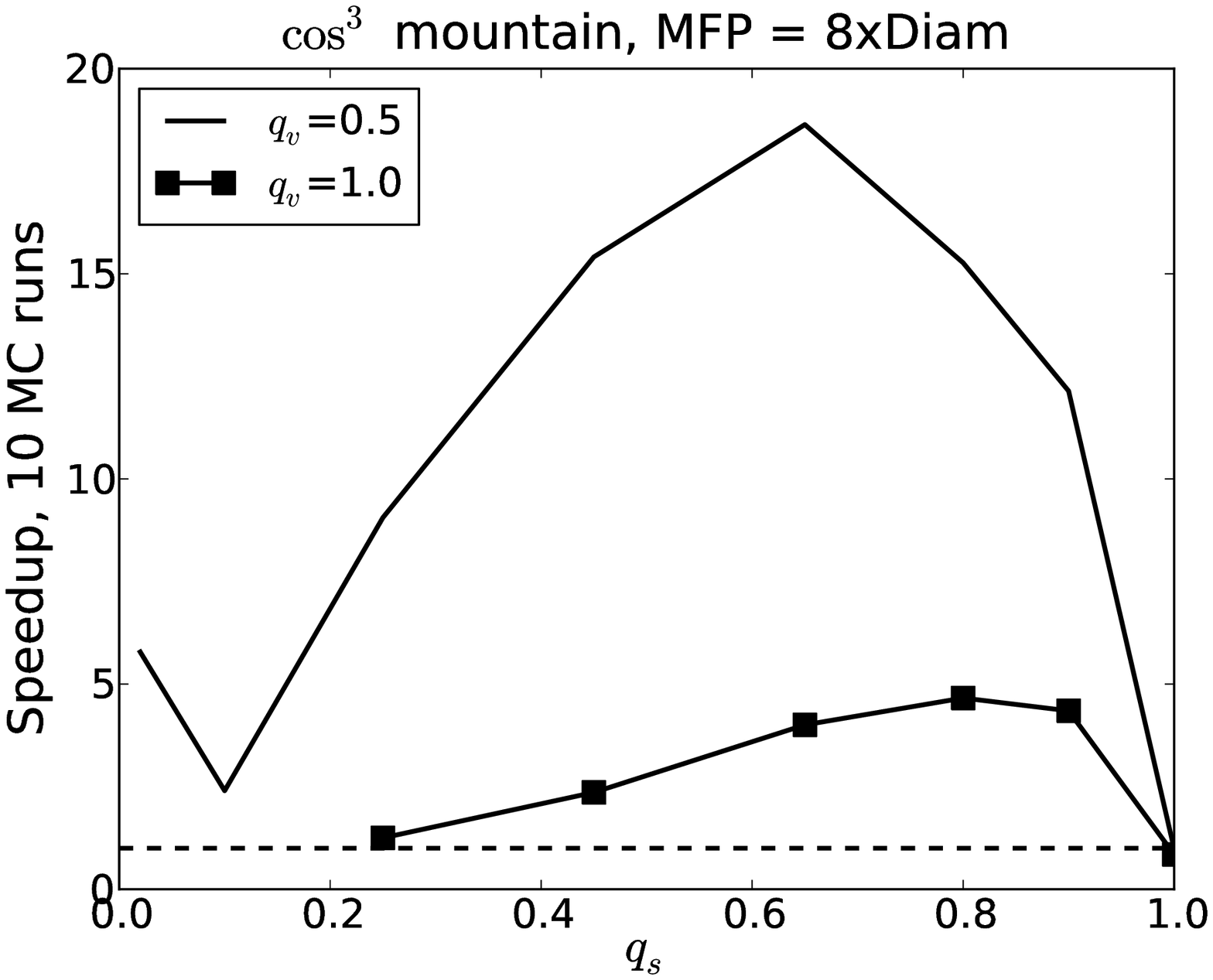}\\
  \includegraphics[width=0.43\textwidth]{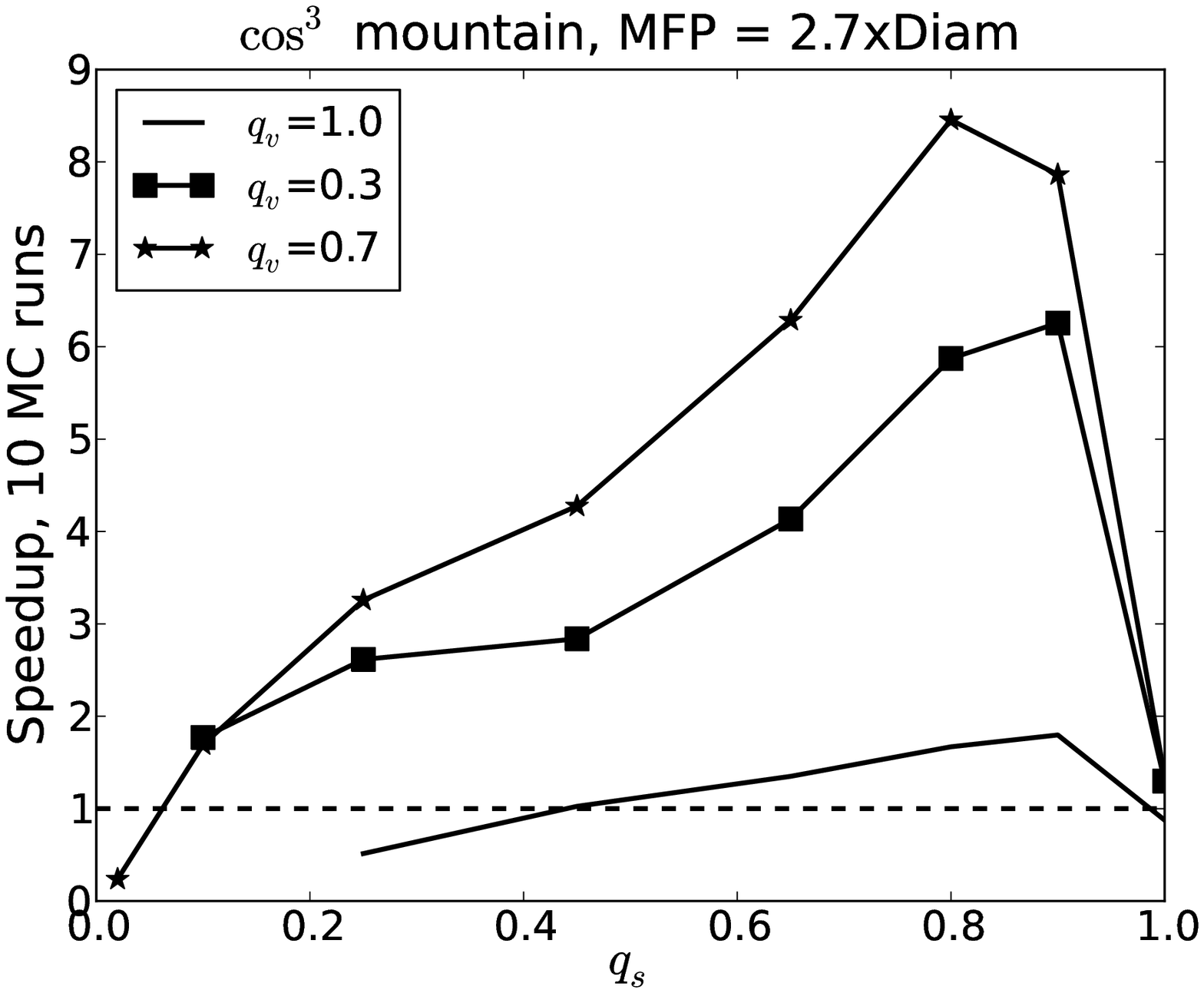}
  \includegraphics[width=0.43\textwidth]{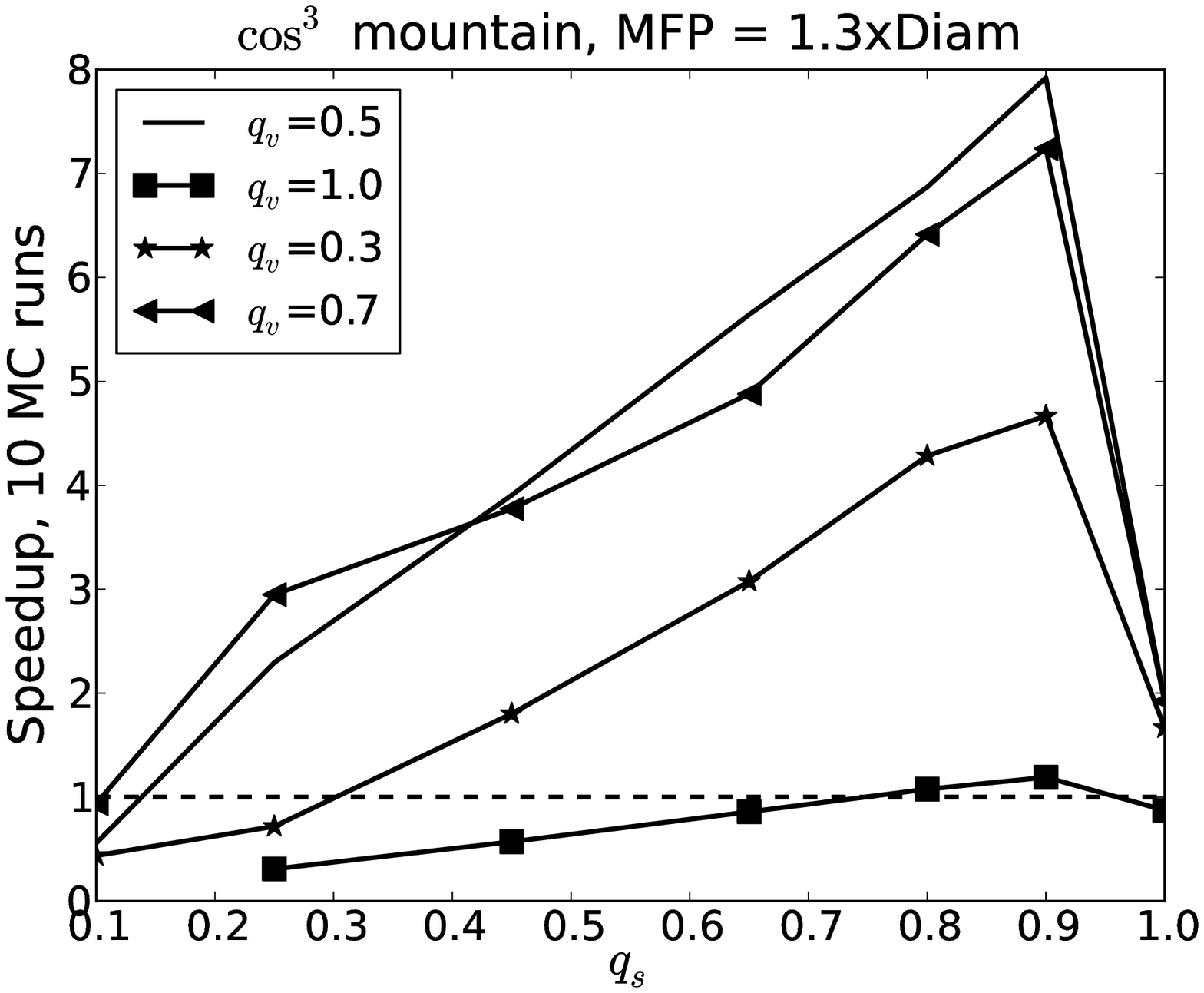}
  \caption{Speedup when using both surface adjoint approximation $\Is$ (with parameter $q_s$) and heuristic atmospheric scattering (with parameter $q_v$)}
  \label{figure:speedup_with_heuristic}
  \end{center}
\end{figure}

Our implementation swept both $q_s$ and $q_v$.  As expected, we see
decreasing speedup with increasing atmospheric scattering strength $\sigma$.  See Fig.~\ref{figure:speedup_with_heuristic}.

It is important to note that use of adjoint-enhanced surface
scattering, and heuristic atmospheric scattering ($q_s<1$, $q_v<1$)
together is especially helpful.  In fact, even with a small
MFP = 1.3$\cdot$Diam (Diam is the maximal diameter of the simulation domain $R$), we realize good speedup when $q_s=0.9$, $q_v<1$.
Note that if either $q_s=1$ or $q_v=1$ (so no use of either SAI or 
heuristic scattering adjustment), speedup almost
disappears.  This is slightly counter-intuitive but may be explained as follows:  Each method (SAI or heuristic) significantly increases the number of paths in two significant classes (surface-only and atmosphere-to-detector).  Therefore, variance from these path-classes is all but eliminated.  Supposing each of these path-classes accounts for $2/5$ of the total paths reaching the detector, by themselves they can only reduce variance by a factor of $1/(1-2/5)=5/3$.  However, together they can reduce variance by a factor of $1/(1-4/5)=5$.

As one can see, selection of the parameters $q_s$ and $q_v$ makes a significant difference in the resultant variance.  We provide some heuristics here and refer the reader to \cite{BalImportanceArxiv} for more details.  When $q_s\to0$ most of the photons will travel on the surface only.  The photons that take a route prescribed by the heuristic chain must then carry an additional weight $=1/q_s$ to compensate for this.  For this reason, picking $q_s$ too small results in increased variance.  A similar argument holds for $q_v$.  That the optimal $q_s$ is so close to $1$ (and greater than the optimal $q_v$) can also be explained by the fact that paths interacting exclusively with the surface are less likely to occur (in the analog world) than those interacting with the surface and atmosphere.

\section{Conclusion and Outlook}
\label{section:conclusion}

A novel method for Monte Carlo transport was presented that uses an approximation of the adjoint (ignoring atmospheric effects) to reduce variance in simulations, equivalently, accelerate convergence to a specified accuracy.  This algorithm, the Surface Adjoint Importance (SAI) method, may be combined with any unbiased method to significantly reduce variance coming from surface interactions when the overlaying atmosphere is optically thin.  If it is combined with a method that reduces variance coming from atmospheric interactions, significant overall variance reduction is achieved.  The implementation is relatively simple, requiring only an approximate adjoint transport solver for the boundary which adds virtually no overhead to the Monte Carlo computation time.  

A possible application of this kind of accelerated Monte Carlo modeling in remote sensing is to address ``adjacency'' effects caused by highly variable terrain, including built environments (urban canyons).  The standard adjacency effect is observed when an aerosol layer of moderate optical thickness mixes in an imaging detector's pixel light that has been reflected off surface elements with contrasting albedos in neighboring pixels.  This is now a solved problem in the case of a variable-but-flat surface under a uniform atmosphere \cite{Lyapustin02}.  However, adjacency effects caused by non-flat terrain are only beginning to be explored, particularly in the thermal IR (where $Q(r,v)$ is determined by temperatures and emissivities).

On a broader scale, our work is an illustration of a modular approach to variance reduction whereby different interactions are handled separately and then pieced together in an unbiased manner.  Specifically, these different interactions could be pieced together as in Algorithm \ref{alg:regularized_sai}.  

For instance one can envision a ``cloud'' module where radiation transport inside the cloud (dominated by multiple scattering) is treated off-line in some judicious approximation, and then incorporated into complex scene simulation.  In applications driven by surface property retrievals from remote sensing data, efficient modularized Monte Carlo modeling would open the door to advanced atmospheric compensation schemes with broken-cloud capability.  This is another wide open frontier recently explored in \cite{BartlettSchott2009}.

%%%%%%%%%%%%%%%%%%%%%%%%%%%%%%%%%%%%%%%%%%%
%%%%%%%%%%%%%%%%%%%%%%%%%%%%%%%%%%%%%%%%%%%
%%%%%%%%%%%%%%%%%%%%%%%%%%%%%%%%%%%%%%%%%%%

\section*{Acknowledgments}

This work was supported in part by DOE/NNSA Grant No. DE-FG52-08NA28779
and NSF Grants Nos. DMS-0804696 and PHY05-51164, as well as NSF Research Training Grant No. DMS-060DMS-0602235.  AD wishes to thank the Kavli Institute for Theoretical Physics at UC Santa Barbara for hospitality and stimulation while finishing this manuscript.  

\appendix
%\begin{appendix}

%\section{Appendix}
\section{Appendix: Numerical solution to the adjoint problem}
\label{section:appendix}

\setcounter{equation}{0}
\numberwithin{equation}{section}

%\subsection{Numerical solution to the adjoint problem}
%\label{subsection:numerical_solution}

Here, at discretization level $h$, we approximate $\Ih\approx\Is$.

To simplify computation of our numerical solution we make the assumption
\begin{align*}
  \PFsurf{r}{v}{v'} &= \one_{\nu_r\cdot v>0}(r,v)\kappa(r,v'), 
\end{align*}
and recall that $g(r,v) = g_0(r)=constant$ whenever $\nu_r\cdot v>0$
so that $g(r,v) = g_0(r)$. The result is that $\Is$ is then a function of position only.  This significantly improves the speed of solving the adjoint problem, as well as the memory requirements for using it.  Theoretical results in this paper do not need this assumption, which we make here as a matter of convenience.

We will now discretize the coefficients and approximate the integral operator appearing on the right hand side of \eqref{eq:psis}, denoted now by $T$.  For $r_1\in\pR$,
\begin{align*}
  T\Is(r_1,v_1) &= \alpha(r_1) \int_{\nu_{r_1}\cdot v_2<0} K(r_1,v_2)\Is(r_+(r_1,v_2),v_2)\dv_2.
\end{align*}
Notice that $T f$ is function depending only on $r$, and in fact only on the boundary values of $f$.  Since $g$ depends only on $r$, $\Is = \sum_{k=0}^\infty T^k g$ will depend only on $r$ and whether or not $\nu_r\cdot v>0$.  We thus define 
\begin{align*}
  \varphi(r) :&= \Is(r,v),\qquad r\in\pR,\,\,\nu_r\cdot v>0.
\end{align*}
We find that $\varphi:\pR\to\Rone$ satisfies the equation
\begin{align*}
  \varphi &= \calA\varphi + g_0,\qquad
  \calA f(r_1) := \alpha(r_1)\int_{\nu_{r_1}\cdot v_2<0} K(r_1,v_2)f(r_+(r_1,v_2))\dv_2.
\end{align*}

In discretizing this operator, and integrals over directions in general, we use the change of variables,
\begin{align}
  \begin{split}
    \int_{\nu_r\cdot v<0}f(r_+(r,v),v)\dv &= \int_\pR f(r',v)\pnuN(r,r')\d\mu(r'),\\
    \pnuN(r,r') &:= \frac{\nu_r\cdot(r'-r)}{|r'-r|^d}.
  \end{split}
  \label{align:boundary_change_of_variables}
\end{align}
The term $\pnuN$ is normal derivative (at $r$) of the free-space Green's function for the Laplacian.  One can show (see, e.g., the section on double-layer potentials in \cite{F-PUP-95}) that for $r,r'\in\pR$, $\nu_r\cdot(r'-r)\lesssim |r'-r|^2$.  Therefore it is in fact an integrable function.  When $d=2$ it is moreover bounded.

We now discretize the operator $\calA$.  First split the boundary into non-overlapping segments $\{\pR_j\}_{j=0}^{N_p-1}$ with $\pR_j$ centered at $r_j$, with length $|\pR_j|\leq h$.  Denote by $Rf$ the (orthogonal) projection of $f$ onto the space of piecewise constant functions (constant on each segment $\pR_j$).  We also think of $Rf$ as a vector in $\Rone^{N_p}$ and $Rf_j$ its components.  Then, after the change of variables \eqref{align:boundary_change_of_variables} we have (at gridpoint $r_i$)
\begin{align}
  \begin{split}
    \calA f(r_i) &= 
    \alpha(r_i)\int_\pR K(r_i,\widehat{r-r_i})\pnuN(r_i,r)f(r)\d\mu(r)\\
    &\approx \alpha(r_i)\sum_{\substack{0\leq j\leq N_p-1\\j\neq
        i}}|\pR_j| K(r_i,\widehat{r_j-r_i})\pnuN(r_i,r_j)f(r_j)
    \\
    &:= \sum_{j}A^h_{ij}Rf_i.
  \end{split}
  \label{align:Q_approx}
\end{align}
This implicitly defines the matrix $A^h$.  

We now define our discrete approximation to $\varphi$ as the piecewise constant function (vector) $\varphih$ solving
\begin{align}
  \varphih &= A^h \varphih + Rg.
  \label{align:varphih_equation}
\end{align}
We then define approximations $\Ih\approx\Is$,
\begin{align}
  \Ih(r,v) :&= \varphih(r), \quad r\in\pR,\,\,\nu_r\cdot v>0.
  \label{align:psih_psioh}
\end{align}
Note that, in our implementation, we have chosen to represent angular integrals as integrals over the boundary. This works for two reasons.  First, as our adjoint solution depends only on position it is convenient to evaluate these sums.  Second, if instead a discretization were chosen that was uniform in angle, then (with only finitely many angles) one would often miss the (small) detector in evaluation of the integral.

%\end{appendix}

%%%%%%%%%%%%%%%%%%%%%%%%%%%%%%%%%%%
%%%%%%%%%%%%%%%%%%%%%%%%%%%%%%%%%%%

%\bibliographystyle{plain}
%\bibliography{bibliography_ian,bibliography} 

\end{document}